\documentclass[final,5p,times,twocolumn]{elsarticle}

\usepackage{lineno,hyperref}
\modulolinenumbers[5]

\biboptions{authoryear}

\begin{document}

\begin{frontmatter}

\title{Systematic biases when using deep neural networks for annotating large catalogs of astronomical images}
\author{Sanchari Dhar, Lior Shamir}
\address{Department of Computer Science, Kansas State University}
\date{}




\begin{abstract}
Deep convolutional neural networks (DCNNs) have become the most common solution for automatic image annotation due to their non-parametric nature, good performance, and their accessibility through libraries such as TensorFlow. Among other fields, DCNNs are also a common approach to the annotation of large astronomical image databases acquired by digital sky surveys. One of the main downsides of DCNNs is the complex non-intuitive rules that make DCNNs act as a ``black box", providing annotations in a manner that is unclear to the user. Therefore, the user is often not able to know what information is used by the DCNNs for the classification. Here we demonstrate that the training of a DCNN is sensitive to the context of the training data such as the location of the objects in the sky. We show that for basic classification of elliptical and spiral galaxies, the sky location of the galaxies used for training affects the behavior of the algorithm, and leads to a small but consistent and statistically significant bias. That bias exhibits itself in the form of cosmological-scale anisotropy in the distribution of basic galaxy morphology. Therefore, while DCNNs are powerful tools for annotating images of extended sources, the construction of training sets for galaxy morphology should take into consideration more aspects than the visual appearance of the object. In any case, catalogs created with deep neural networks that exhibit signs of cosmological anisotropy should be interpreted with the possibility of consistent bias.

\end{abstract}


\end{frontmatter}


\section{Introduction}
\label{introduction}

In the past two decades, autonomous digital sky surveys powered by robotic telescopes have been becoming increasingly important in astronomy, and have been revolutionizing astronomy research. The ability to collect very large databases of astronomical images and make them accessible through on-line services increased the productivity of telescope systems, leading to unprecedented discovery power \citep{edwards2014astronomy}. It also allows more potential contributors to participate in astronomy research, as the accessibility of the data provides more researchers with access to high-quality data without the need to own and operate highly expensive research facilities.

Perhaps the first major comprehensive autonomous digital sky survey is the Sloan Digital Sky Survey \citep{york2000sloan}, with considerable success and revolutionary impact on astronomy. The overwhelming success of SDSS was followed by other powerful sky surveys such as the Panoramic Survey Telescope and Rapid Response System \citep{kaiser2010pan} and the Dark Energy Survey \citep{abbott2016dark}. Future ventures such as the Vera Rubin Observatory and the space-based Euclid mission will provide even more powerful imaging capabilities, leading to far greater databases and consequently discovery power.

As digital sky surveys image hundreds of millions and even billions of astronomical objects, it is clear that manual analysis of the data is impractical. One of the most challenging tasks in the analysis of the image data acquired by digital sky surveys is the morphological analysis of extended objects. Unlike point sources, extended objects can have complex morphology, and their analysis require sophisticated computational methods. Tasks related to automatic annotations of galaxies can include broad classification of galaxies to elliptical or spiral \citep{gonzalez2018galaxy,ibrahim2018galaxy,shamir2009automatic,banerji2010galaxy,gauci2010machine,jimenez2020galaxy}, or annotation of a more comprehensive set of morphological descriptors of galaxies \citep{kuminski2014combining,dieleman2015rotation,diaz2019classifying,zhu2019galaxy}. Other tasks can include automatic detection of rare galaxies \citep{shamir2012automatic,timmis2017catalog,jacobs2019extended,jacobs2019finding,davies2019using}, spiral arm segments \citep{davis2014sparcfire}, unsupervised analysis of galaxy morphology \citep{schutter2015galaxy}, or separating galaxies from stars \citep{sevilla2018star}.

In the past decade, deep convolutional neural networks (DCNNs) have been becoming increasingly more common in machine vision. DCNNs provide good performance without the need to design specific algorithms for each image analysis problem. The availability of open source libraries makes DCNNs available also for those who are not necessarily machine learning experts. As they become popular in almost all fields that involve machine vision, DCNNs have also become common in astronomy. Among other tasks, they are also used for the purpose of automatic annotation of galaxy images \citep{dieleman2015rotation,cheng2020optimizing,gonzalez2018galaxy,barchi2020machine,dominguez2018improving,khan2019deep}. Due to their efficiency and speed, DCNNs are currently the most common solution for the annotation of very large datasets of galaxy images.

However, while DCNNs have the important advantages mentioned above, they also have several weaknesses. One of the main downsides of DCNNs is the ``black box" nature of its classification process. DCNNs are trained by data, and the features are determined automatically during training to optimize the classification accuracy. These features are normally complex and non-intuitive, making it difficult to conceptualize the way the classifications are being made by the neural network. For that reason, the performance of DCNNs is normally determined empirically through a quantitative analysis using the test data, rather than a qualitative analysis that is based on how the data are actually being processed by the DCNNs. Since it is difficult to define what the DCNNs ``learn" from the data, such systems should be used with caution \citep{lapuschkin2019unmasking,schramowski2020making}.

Here we demonstrate that a DCNN system used through a typical supervised machine learning process to identify galaxy morphology can provide good accuracy in the classification of the galaxies, but at the same time can have subtle but consistent bias. When applied to large datasets, that bias can lead to consistently biased catalogs. If the bias is ignored in consequent analysis of the catalog, it can lead to observations that do not reflect the distribution of the data in the real sky.

\section{Method}
\label{method}

We train a neural network to classify between elliptical and spiral galaxies. The distribution of elliptical and spiral galaxies in one part of the sky are expected to be statistically the same as the distribution of elliptical and spiral galaxies in other parts of the sky. That is, when the dataset of galaxies is large, the shape of elliptical galaxies observed in a certain RA and declination range is expected to be the same as the shape of elliptical galaxies in any other RA and declination range. An expert observing an image of a random elliptical galaxy, and given no other information about the object, will not be able to make a knowledgeable guess of the RA and declination of that galaxy.

According to the null hypothesis, the neural network is trained by the morphology of the galaxies, and therefore the classification output of the neural network depends only on the morphology of the galaxy. In that case, if the deep neural network is trained with a high number of galaxies, it will perform the same way (within statistical error) regardless of the location of the test galaxy in the sky. Otherwise, the neural network is sensitive also to the location of the galaxy in the sky, and therefore can lead to a certain bias. Even if such bias is small, when annotating a very large number of galaxies that bias can be statistically significant. For instance, if in a certain part of the sky the neural network tends to classify more galaxies as spiral, a catalog generated by that network will show cosmological anisotropy such that a certain direction of observation has more spiral galaxies compared to other directions of observation.

In this study we train a deep neural network to classify between elliptical and spiral galaxies by using training galaxies imaged in the same part of the sky. We then compared the results using the same test set, but such that the training spiral galaxies are taken from one part of the sky, and the test spiral galaxies are taken from another part of the sky. If the neural network classifies galaxies just by their morphology, both neural networks should provide the same confusion matrix, within statistical error. However, if the confusion matrices are different, it means that the neural network also learns differences in the sky background, and can be affected based on the location of the galaxy in the sky. When applied to a large dataset, such behavior of the neural network can lead to differences in the distribution of the annotations based in different parts of the sky. That can consequently lead to slightly but consistently biased data products.



\subsection{Deep convolutional neural network}
\label{cnn}

The deep convolutional neural network used in this experiment is an expansion of the common LeNet-5 architecture \citep{lecun1998gradient}, implemented using the Keras library \citep{chollet2015keras,chollet2018keras}, and adjusted to the input size of images with dimensionality of 120$\times$120 pixels. The deep neural network model is based on the sequential model of Keras with five convolution layers and four max-pooling layers. 
These layers are followed by flattening and fully connected layers. 

\begin{figure}[h!]
\centering
\includegraphics[scale=0.65]{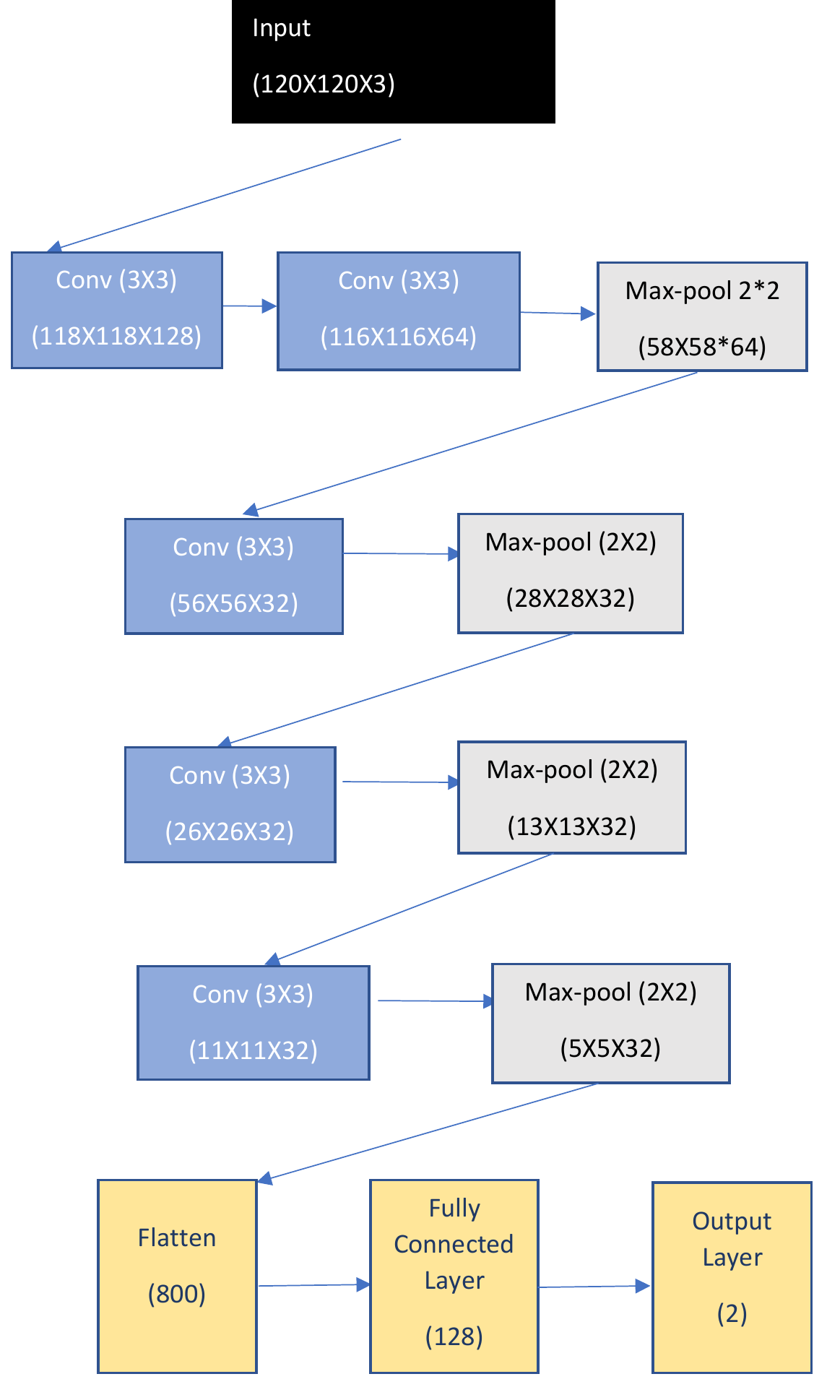}
\caption{Diagram of the structure of the convolutional neural network.}
\label{fig:diagram}
\end{figure}

The activation function used in most layers of the convolutional neural network is Rectified Linear Unit (ReLU), except for the output layer, where we use the sigmoid activation function. Figure~\ref{fig:diagram} shows the diagram of the structure of the convolutional neural network. During compilation the model uses the Adam (Adaptive Moment estimation) optimizer \citep{kingma2014adam}, with an adaptive learning rate and the binary cross entropy is used as the loss function because of binary classification. The network was trained with 250 epochs, similarly to the training of the neural network described in \citep{goddard2020catalog}, which was sufficient to reach the peak accuracy.

\section{Data}
\label{data}

Datasets from two major digital sky surveys were used - SDSS and Pan-STARRS. To  have training and test sets of spiral galaxies, we used catalogs of galaxies annotated by their broad morphology. The image data in both datasets are the 120$\times$120 JPG images downloaded by using the {\it cutout} service. For SDSS, the annotation of spiral and elliptical galaxies were taken from a catalog of galaxies annotated by their broad morphology \citep{kuminski2016computer,paul2018catalog}. Each galaxy in the catalog is provided with its annotation, and the certainty of the annotation in the range (0.5,1), where 1 is the maximum certainty for the galaxy to belong in the morphological type it is annotated. To ensure the accuracy of the annotations, only galaxies with certainty threshold of 0.9 or higher were used. These galaxies have certainty of their annotation of $\sim$98\% compared to the ``superclean" Galaxy Zoo annotations \citep{kuminski2016computer}. The total number of galaxies in the catalog is $\sim2.9\cdot10^6$.

\begin{figure}
\includegraphics[scale=0.70]{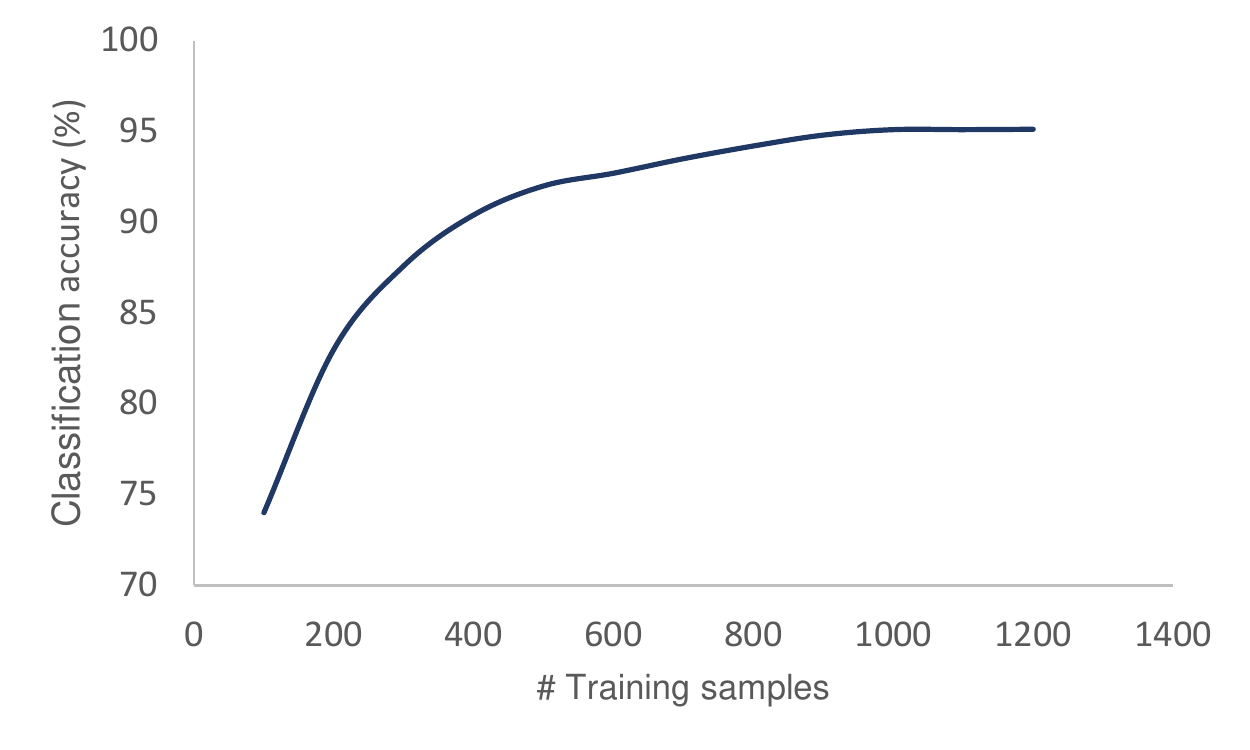}
\caption{The classification accuracy of the convolutional neural network when applied to the Pan-STARRS galaxy images. The classification accuracy increases as the training set gets larger, until it reaches $\sim$1,000 traning images per class.}
\label{accuracy}
\end{figure}

A similar catalog was also used for the Pan-STARRS data. The catalog of Pan-STARRS galaxies contained $\sim1.7\cdot10^6$ automatically annotated galaxies imaged by Pan-STARRS \citep{goddard2020catalog}. Like with the SDSS galaxies, only galaxies with annotation certainty of 90\% or higher were used. While that threshold limit reduced the size of the training set, it ensured that the dataset for training and testing the convolutional neural network is clean, as shown by comparison of the threshold of the classification algorithm to manual classification of the galaxies \citep{goddard2020catalog}. Figure~\ref{accuracy} shows the classification accuracy of the convolutional neural network described in Section~\ref{cnn} and applied to the Pan-STARRS galaxy images, such that 8,000 images per class are used for testing, and the size of the training set varies. As the figure shows, the classification accuracy improves until the training set reaches the size of $\sim$1,000 images per class, after which it stabilizes. 

For Pan-STARRS data, datasets of galaxies were taken from two opposite hemispheres in the sky. The RA ranges of the sky regions used in the experiments are $(0^o-20^o)$, and $(180^o-200^o)$. The declination in both cases is $(0^o-20^o)$. These regions were chosen for being far from each other in the sky, but also because they contain a sufficient number of galaxies. Additionally, a dataset of Pan-STARRS galaxies in the RA range $(180^o-200^o)$ and declination range of $(20^o-40^o)$ is also used. The sky regions used in the SDSS data are RA ranges $(240^o-260^o)$ and $(15^o-35^o)$ for the RA. The declination range is $(-10^o-40^o)$. Table~\ref{data_pansStarrs_SDSS} summarizes the number of objects taken from each sky region in each sky survey, and the number of objects used for testing and training. Naturally, the training and test sets are completely orthogonal.

\begin{table*}[h]
\scriptsize
\begin{tabular}{|c|c|c|c|c|c|c|c|} 
 \hline
Dataset     & Sky     & RA     & Dec   & Train      & Train   & Test         & Test  \\ 
             & survey & range  & range & Elliptical &  spiral &   elliptical &  spiral \\ 
\hline
 1 & Pan-STARRS & 0$^o$-20$^o$ & 0$^o$-20$^o$ & 3000 & 3000 & 8000 & 8000 \\
 2 & Pan-STARRS & 180$^o$-200$^o$ & 0$^o$-20$^o$ & 3000 & 3000 & 8000 & 8000\\ 
 3 & Pan-STARRS & 180$^o$-200$^o$ & 20$^o$-40$^o$& 3000 & 3000 & 8000 & 8000\\ 
 A & SDSS & 240$^o$-260$^o$ & -10$^o$- 40$^o$ & 2000 & 2000 & 3000 & 3000\\
 B & SDSS & 15$^o$-35$^o$ & -10$^o$- 40$^o$ & 2000 & 2000 & 3000 & 3000 \\ 
 \hline
\end{tabular}
\caption{The datasets used in the experiment, and the number of training and test galaxies in each dataset. RA and declination ranges are in degrees.}
\label{data_pansStarrs_SDSS}
\end{table*}

In most machine learning systems most samples are allocated for training, in order to improve the accuracy of the classifier. Here, however, most samples are allocated for testing. As shown in Figure~\ref{accuracy}, increasing the number of training samples is not expected to increase the accuracy of the neural network. The smaller number of training samples leaves a higher number of test samples, allowing better statistical analysis of possible biases in the neural network classifier as will be discussed in Section~\ref{results}.

Table~\ref{field_differences} shows several certain properties of the different fields from which the galaxy images where acquired. The information is taken from Pan-STARRS DR1 and SDSS DR8, which are the data releases from which the images were taken. In SDSS, the type of objects was determined by using the SDSS classification system. Pan-STARRS DR1 does not have galaxy/star classification, and therefore that was done by considering stars as objects that their PSF {\it i} magnitude subtracted by their Kron {\it i} magnitude was greater than 0.05, as was done in \citep{timmis2017catalog}. The table shows a substantial difference in star population between dataset A and B, while Datasets 1 and 2 are relatively consistent in terms of population of stars, galaxies, and average magnitudes.

\begin{table}[h]
\scriptsize
\begin{tabular}{|c|c|c|c|c|c|} 
 \hline
Dataset     & \#     & \#       & Mean star   & Mean galaxy      & $\sigma$ galaxy    \\ 
             & stars & galaxies  & magnitude & magnitude          &  magnitude       \\ 
\hline
 1 &  13,119,580 & 9,118,961 & 22.768 & 21.567 & 2.01  \\
 2 & 13,522,422 & 9,134,103 & 22.712 & 21.771 & 2.05 \\ 
 3 & 11,714,198 & 7,995,078 & 22.747 & 21.934 & 2.11 \\ 
 A & 16,049,710 &  10,561,930 & 21.0437 & 21.727 & 3.387  \\
 B & 8,046,344 & 11,752,972 & 21.3569 & 21.5327 & 3.229  \\ 
 \hline
\end{tabular}
\caption{Properties of the different fields from which the different datasets where acquired.}
\label{field_differences}
\end{table}

\section{Results}
\label{results}

The DCNN method described in Section~\ref{cnn} was tested with different combinations of the datasets described in Section~\ref{data}. For the baseline experiment we trained and tested the model with the galaxies in the same right ascension and declination ranges. Table~\ref{basecase1} shows the confusion matrix when classifying the Pan-STARSS galaxies using the spiral and elliptical galaxies from Dataset 1 for both training and test purpose. The number of training and test galaxies are specified in Table~\ref{data_pansStarrs_SDSS}. 

\begin{table}[h]
\begin{tabular}{ |c|c|c| } 
 \hline
       & Elliptical & Spiral \\ 
\hline
 Elliptical & 7850 & 150   \\ 
 Spiral     & 756  & 7244  \\ 
 \hline
\end{tabular}
\caption{Confusion matrix of the classifications when using Dataset 1 for both training and testing. The rows are the ground truth of the galaxy morphological types, which total at 8000 elliptical galaxies and 8000 spiral galaxies. The columns show how the galaxies were classified by the CNN.}
\label{basecase1}
\end{table}

As the table shows, more spiral galaxies were incorrectly classified as elliptical galaxies compared to elliptical galaxies classified incorrectly as spiral. Therefore, if applied to a very large dataset of galaxies, it is expected to show a slightly higher fraction of elliptical galaxies than spiral galaxies. However, a slight bias is expected from any classifier, and such bias is not necessarily expected to lead to an observation of a large-scale anisotropy. The reason a bias in the classifier is not expected to lead to an observation of anisotropy is that such bias is expected in all parts of the sky. That is, regardless of the location of the test galaxies in the sky, a higher number of elliptical galaxies is expected, and therefore the ratio between elliptical and spiral galaxies is not expected to change throughout the sky. 

The number of elliptical and spiral galaxies is not necessarily equal, and the separation between a spiral and an elliptical galaxy is not necessarily a pure binary classification problem due to in-between cases. Therefore, a slight but consistent bias of the algorithm might not necessarily lead to false large-scale anisotropy. Certainly, the performance of such algorithms also depends on the imaging, as higher resolution images allow the identification of spiral features of galaxies, increasing the number of spiral galaxies compared to elliptical galaxies \citep{goddard2020catalog}. If spiral features are identified, that indicates that the galaxy is indeed spiral. However, if spiral features are not identified it could also be because the image resolution does not allow the identification of the spirality of the galaxy \citep{goddard2020catalog}.  

The same analysis was also done using Dataset 2. Table~\ref{basecase2} shows the confusion matrix when classifying the Pan-STARSS galaxies using Dataset 2 for both training and testing.  Table~\ref{basecase_combined} shows the confusion matrix when Datasets 1, 2, and 3 are combined into a single dataset with 18,000 training images and 48,000 test images. 

\begin{table}[h]
\begin{tabular}{ |c|c|c| } 
 \hline
       & Elliptical & Spiral \\ 
\hline
 Elliptical & 7699 & 301  \\ 
 Spiral     & 450 & 7550  \\ 
 \hline
\end{tabular}
\caption{Confusion matrix of the classification when using Dataset 2 for both training and testing.}
\label{basecase2}
\end{table}

\begin{table}[h]
\begin{tabular}{ |c|c|c| } 
 \hline
       & Elliptical & Spiral \\ 
\hline
 Elliptical & 23299 & 751  \\ 
 Spiral     & 1646 & 22304  \\ 
 \hline
\end{tabular}
\caption{Confusion matrix of the classification when combining Datasets 1, 2, and 3 into a single dataset.}
\label{basecase_combined}
\end{table}

The error rate for the spiral and elliptical galaxies is somewhat different from the error rate shown in Table~\ref{basecase1}, and the difference provides a certain indication of a link between the performance of the classifier and the part of the sky from which the galaxies are taken. 
However, in these two experiments the training and test data are different, and therefore no conclusive evidence of a link between a consistent bias of the classifier and the part of the sky from which the training data are taken can be inferred.


To further investigate a possible link between the part of the sky from which the training data are taken and a consistent bias of the classifier, the training set was designed such that the spiral galaxies were taken from Dataset 1, and the elliptical galaxies from Dataset 2. The test galaxies are all from Dataset 1. Table~\ref{confusion_matrix_group2group1_group1_pan} shows the confusion matrix of the predictions made by the classifier.

\begin{table}[h]
\begin{tabular}{ |c|c|c| } 
 \hline
       & Elliptical & Spiral \\ 
\hline
 Elliptical & 7749 & 251   \\ 
 Spiral     & 408  & 7592  \\ 
 \hline
\end{tabular}
\caption{Confusion matrix of the classifications when using spiral galaxies from Dataset 1, elliptical galaxies Dataset 2 for training, and all test galaxies from Dataset 1 for testing.}
\label{confusion_matrix_group2group1_group1_pan}
\end{table}

As the table shows, although the test galaxies are the same, the distribution of the galaxies is different from the results of the experiment shown in Table~\ref{basecase1}. According to Table~\ref{confusion_matrix_group2group1_group1_pan}, more spiral galaxies are classified correctly, while more elliptical galaxies are classified incorrectly.

According to Table~\ref{confusion_matrix_group2group1_group1_pan}, $\sim$49.02\% are predicted as spiral, and therefore the probability of a galaxy to be predicted as spiral in that dataset is 0.4902. According the results of Table~\ref{basecase1}, merely 7,394 galaxies were classified as spiral. According to binomial distribution, if the success probability is 0.4902, the probability of having 7,394 or less successful events is $P<10^{-5}$. That shows that although the test set is identical, the predictions are significantly different when using training data from different parts of the sky.

The high probability shows that if the classifier was applied to galaxy images that are not labeled with ground truth, it would have shown a statistically significant difference between the frequency of spiral galaxies in the sky region of $(0^o<\alpha<20^o,0^o<\delta<20^o)$ and the frequency of spiral galaxies in the sky region $(180^o<\alpha<200^o,0^o<\delta<20^o)$. These differences could be interpreted as evidence for cosmological-scale anisotropy.

Differences in the confusion matrices can also be due to fluctuations in the classification process. DCNNs are trained with the data, and therefore different training data is expected to lead to different classification results. To test these fluctuations empirically, the experiment shown in Table~\ref{confusion_matrix_group2group1_group1_pan} was repeated with 100-fold bootstrapping, such that in each run the galaxies were separated randomly to training and test sets. The results of the 100 runs show a mean of 247.4$\pm$1.1 misclassified elliptical galaxies, and a mean of 406.1$\pm$1.4 misclassified spiral galaxies. The maximum number of misclassified elliptical galaxies was 291, which is smaller than the 756 misclassified spiral galaxies in the confusion matrix of Table~\ref{basecase1}.

To further examine a possible link between the selection of the training data and the behavior of the classifier, a similar experiment was performed such that the training data was spiral galaxies from Dataset 1 and elliptical galaxies from Dataset 2. Then, the performance of the classifier was tested with test data from Dataset 2. 
Table~\ref{confusion_matrix_group2group1_group2_pan} shows the confusion matrix of the experiment.

\begin{table}[h]
\begin{tabular}{ |c|c|c| } 
 \hline
       & Elliptical & Spiral \\ 
\hline
 Elliptical & 7791 & 209 \\ 
 Spiral     & 462  & 7538  \\ 
 \hline
\end{tabular}
\caption{Confusion matrix of the classifications when using spiral galaxies from Dataset 1, elliptical galaxies from Dataset 2 for training, and Dataset 2 used for testing.}
\label{confusion_matrix_group2group1_group2_pan}
\end{table}

As the table shows, when comparing to the classifications of the same set of galaxies with a neural network that was trained with galaxies from the same sky region as the test data, the number of misclassified spiral galaxies increases from 450 to 462, and the number of misclassified elliptical galaxies decreases from 301 to 209. If the probability of misclassified elliptical galaxy is $\frac{301}{8000}=0.037625$, the probability of having 209 or less misclassified elliptical galaxies from the 8,000 galaxies that were classified is $(P<10^{-5})$. Because the galaxies have ground truth we can be certain that the reason for the difference is not of astronomical origin, but higher similarity of the test elliptical galaxies from Dataset 2 and the training elliptical galaxies from Dataset 2, which leads to a consistent bias in the classification.

In an empirical test the dataset was separated randomly into training and test sets of the same sizes as specified in Table~\ref{data_pansStarrs_SDSS}, but the experiment was repeated 100 times such that in each run the galaxy images were randomly allocated to training and test sets. After 100 runs, the mean number of misclassified elliptical galaxies was 216.2$\pm$1, and the mean misclassified spiral galaxies was 473$\pm$1.4. In 100 runs the maximum number of misclassified elliptical galaxies was 277, which is below the 301 misclassified elliptical galaxies shown in the confusion matrix of Table~\ref{basecase2}.



In another experiment, the training set contained spiral galaxies from Dataset 2, and elliptical galaxies from Dataset 1. Tables~\ref{confusion_matrix_g1_g2} and~\ref{confusion_matrix_g2_g1} show the confusion matrix when applying the classifier to the test samples of Dataset 1 and Dataset 2, respectively.

\begin{table}[h]
\begin{tabular}{ |c|c|c| } 
 \hline
       & Elliptical & Spiral \\ 
\hline
 Elliptical & 7592 & 408\\ 
 Spiral     & 551  & 7449  \\ 
 \hline
\end{tabular}
\caption{Confusion matrix of the classification when training the neural network with spiral galaxies from Dataset 2, elliptical galaxies from Dataset 1, and  testing the classifier with spiral and elliptical galaxies from Dataset 1.}
\label{confusion_matrix_g1_g2}
\end{table}

\begin{table}[h]
\begin{tabular}{ |c|c|c| } 
 \hline
       & Elliptical & Spiral \\ 
\hline
 Elliptical & 7514 & 486 \\ 
 Spiral     & 342  & 7658  \\ 
 \hline
\end{tabular}
\caption{Confusion matrix of the classification when the spiral galaxies of the training set are from Dataset 2, the elliptical galaxies of the training set are from Dataset 1, and the test samples are the test galaxies from Dataset 2.}
\label{confusion_matrix_g2_g1}
\end{table}

The results show that although the training set is the same in both cases, each dataset provided different results. For the test samples of Dataset 1, a higher number of spiral galaxies was misclassified as elliptical galaxies, while when classifying the test samples of Dataset 2 more elliptical galaxies were classified as spiral galaxies. That is, the classifier showed 7,857 spiral galaxies in Dataset 1, and 8,144 spiral galaxies in Dataset 2. Because the galaxies are annotated with ground truth, we can conclude that the reason for the difference is not an actual higher number of spiral galaxies in the real sky at $(180^o<\alpha<200^o,0^o<\delta<20^o)$, but a higher similarity of the galaxies in the test set to the galaxies in the training set that were taken from the same part of the sky. However, if the galaxies did not have ground truth, the difference could have been interpreted as an indication of cosmological-scale anisotropy.


\subsection{Difference between close sky regions}

The experiments above tested for the impact when the different classes in the training set are imaged in opposite hemispheres. To test whether the same bias also occurs when the classes of the training data are acquired in closer regions, another experiment was done such that the difference sky regions are neighboring.

Table~\ref{confusion_matrix_g2_g3} shows the confusion matrix when the neural network was trained with elliptical galaxies from Dataset 2, and spiral galaxies from Dataset 3. Dataset 2 was used for testing. These results are compared to the confusion matrix of Table~\ref{confusion_matrix_g3_g2}, showing the classifications when the same test set was used, but the neural network was trained with elliptical galaxies from Dataset 3, and spiral galaxies from Dataset 2. The two confusion matrices show some differences in the classifications, but less substantial than when the training sets were taken from opposite hemispheres. For instance, the difference in the number of misclassified spiral galaxies is about 14\%. That is smaller difference compared to the 62\% difference in misclassified spiral galaxies observed when the training and test sets are from two different hemispheres.

\begin{table}[h]
\begin{tabular}{ |c|c|c| } 
 \hline
       & Elliptical & Spiral \\ 
\hline
 Elliptical & 7755 & 245  \\ 
 Spiral     & 600  & 7400  \\ 
 \hline
\end{tabular}
\caption{The confusion matrix when the classifier is trained with elliptical galaxies from Dataset 2, and spiral galaxies from Dataset 3. The test set is the test galaxies from Dataset 2.}
\label{confusion_matrix_g2_g3}
\end{table}

\begin{table}[h]
\begin{tabular}{ |c|c|c| } 
 \hline
       & Elliptical & Spiral \\ 
\hline
 Elliptical & 7702 & 298  \\ 
 Spiral     & 516  & 7484  \\ 
 \hline
\end{tabular}
\caption{The confusion matrix when the neural network is trained with elliptical galaxies from Dataset 3, and spiral galaxies from Dataset 2. The test set is the test galaxies from Dataset 2.}
\label{confusion_matrix_g3_g2}
\end{table}




\subsection{Experiments with SDSS data}
\label{SDSS}

In addition to the Pan-STARRS data, we also tested data from the Sloan Digital Sky Survey (SDSS). Table~\ref{confusion_matrix2_gA_gB_gA_SDSS} shows the confusion matrix of the classifications of the test galaxies of Dataset A, when the classifier was trained with training spiral galaxies from Dataset A, and training  elliptical galaxies from Dataset B. As before, no galaxies were included in both the training and test sets.  Table~\ref{confusion_matrix2_gA_gB_gB_SDSS} shows the confusion matrix when using the same training set as was used for the experiment shown in Table~\ref{confusion_matrix2_gA_gB_gA_SDSS}.


\begin{table}[h]
\begin{tabular}{ |c|c|c| } 
 \hline
       & Elliptical & Spiral \\ 
\hline
 Elliptical & 2704 & 296 \\     
 Spiral     & 31   & 2969  \\  
 \hline
\end{tabular}
\caption{Confusion matrix of the classification of SDSS dataset when training the convolutional neural network with spiral galaxies from Dataset A and elliptical galaxies from Dataset B. The test samples are from Dataset A.}
\label{confusion_matrix2_gA_gB_gA_SDSS}
\end{table}

\begin{table}[h]
\begin{tabular}{ |c|c|c| } 
 \hline
       & Elliptical & Spiral \\ 
\hline
 Elliptical & 2891 & 109 \\ 
 Spiral     & 85   & 2915  \\ 
 \hline
\end{tabular}
\caption{Confusion matrix of the classification of SDSS galaxies when training the neural network with spiral galaxies from Dataset A and elliptical galaxies from Dataset B, and test set is the test samples of Dataset B.}
\label{confusion_matrix2_gA_gB_gB_SDSS}
\end{table}

As the table shows, the number of misclassified spiral galaxies increases to 85, while the number of misclassified elliptical galaxies drops to 109. If the probability of an elliptical galaxy to be misclassified as spiral galaxy is $\sim$0.0987 as shown in Table~\ref{confusion_matrix2_gA_gB_gA_SDSS}, the binomial distribution probability to have 109 or less misclassified elliptical galaxies is $P<10^{-5}$. 

Like with the Pan-STARRS data, the analysis shows that the classification exhibit different ratio of elliptical and spiral galaxies even when the test set is identical. Because the test samples are exactly the same, and the algorithm is the same, the only possible explanation for the difference is the use of a different training set. For example, the increase in the number of galaxies classified as elliptical can be linked to the fact that the neural network was trained with elliptical galaxies imaged in the same part of the sky of the test samples, while the training spiral galaxies were imaged in a different part of the sky. If the neural network also ``learns'' the sky background, the similarity between the background of the images in the test set and the background of the images in just one of the classes in the training set can lead to bias towards that class. The bias is statistically significant. Therefore, using the same neural network to classify galaxies in the entire sky would lead to a statistically significant difference between the frequency of spiral galaxies in different parts of the sky, what might provide evidence of cosmological-scale anisotropy.

A similar experiment was done such that the training set was made of spiral galaxies from Dataset B and elliptical galaxies from Dataset A. Tables~\ref{confusion_matrix_g1_g1_SDSS} and~\ref{confusion_matrix_g1_g2_SDSS} show the confusion matrices when testing the neural network with test data from Dataset A and Dataset B, respectively. As the confusion matrices show, the bias identified in the results is consistent with the results of the previous experiment.

Although the classifier is the same classifier trained with the same data, the misclassifications of Dataset A are completely different than the misclassifications of Dataset B. When testing the classifier with the test data of Dataset A, much more galaxies are classified as elliptical compared to the confusion matrix produced when the classifier was tested with Dataset B. Given that the probability of a test spiral galaxy in Dataset A to be misclassified as an elliptical galaxy is 0.165, the probability to have 154 misclassified galaxies or less is $P<10^{-5}$.

\begin{table}[h]
\begin{tabular}{ |c|c|c| } 
 \hline
       & Elliptical & Spiral \\ 
\hline
 Elliptical & 2933 & 67 \\ 
 Spiral     & 495  & 2505  \\ 
 \hline
\end{tabular}
\caption{Confusion matrix of the classifications of SDSS galaxies when training the neural network with spiral galaxies from Dataset B and elliptical galaxies from Dataset A. The test set is spiral and elliptical galaxies from the test samples of Dataset A.}
\label{confusion_matrix_g1_g1_SDSS}
\end{table}

\begin{table}[h]
\begin{tabular}{ |c|c|c| } 
 \hline
       & Elliptical & Spiral \\ 
\hline
 Elliptical & 2837 & 163   \\ 
 Spiral     & 154  & 2846  \\ 
 \hline
\end{tabular}
\caption{Confusion matrix of the classifications of SDSS galaxies when training the neural network with spiral galaxies from Dataset B and elliptical galaxies from Dataset A. The test samples are the test spiral and elliptical galaxies from Dataset B.}
\label{confusion_matrix_g1_g2_SDSS}
\end{table}

The results of all experiments are summarized in Table~\ref{summary_table}.

\begin{table}[h]
\footnotesize
\begin{tabular}{ |c|c|c|c|c|c| } 
 \hline
Training     & Training  & Test  & Test         & p(spiral)   &  p(elliptical)  \\ 
spiral         & elliptical    & spiral & elliptical    &              &                     \\
\hline   
 1             & 1               & 1      &      1        &  0.46       &   0.54      \\ 
 2             &   2             &  2     &      2        &  0.49       &    0.51     \\ 
 1+2+3     & 1+2+3   & 1+2+3    & 1+2+3    &  0.48        & 0.52 \\
 1            &    2         &    1        &     1       &  0.49        &  0.51  \\
  1          &      2        &   2         &     2       &  0.48        & 0.52  \\
  2          &      1        &   1        &     1       &  0.49         &  0.51 \\
  2           &     1        &   2        &     2       &  0.51        &   0.49 \\
  3            &   2       &     2        &     2       & 0.48       & 0.52 \\
  2            &  3        &    2         &    2       &  0.49 & 0.51 \\
  A   &    B   &  A  &  A  & 0.55  & 0.45 \\
A & B & B & B &  0.5 & 0.5 \\
B & A & A & A & 0.43 & 0.57 \\
B & A & B & B &  0.5 & 0.5 \\
 \hline
\end{tabular}
\caption{A summary of the experiments shown in this section. The numbers of the datasets correspond to the dataset numbers in Table~\ref{data_pansStarrs_SDSS}.}
\label{summary_table}
\end{table}

\section{Conclusion}
\label{conclusion}

As autonomous digital sky surveys generate vast pipelines of image data, including billions of extended objects with complex morphology, a solid approach to analyze the morphology of these objects is by applying deep convolutional neural networks. For the purpose of supervised machine learning, these networks are trained automatically with labeled ``ground truth" samples, and can then annotate any given new data based on the rules deduced in the training stage.

The application of convolutional neural networks to large image databases collected by digital sky surveys can produce large catalogs of annotated objects.  For instance, the identification of a large number of gravitational lens candidate \citep{jacobs2019extended,jacobs2019finding} would have been impractical without using automation. It is therefore expected that data products generated by deep neural networks will be used by other researchers to answer questions that were difficult to address observationally in the pre-information era, such as the large-scale structure of the universe.

However, while deep neural networks can provide fast annotation with high level of accuracy, they are based on complex and non-intuitive data-driven rules that are difficult to interpret and fully understand. Therefore, these rules can reflect not just the morphology of the galaxy, but in fact any piece of information by which the neural network can differentiate between the different classes of images it is trained with. 

Here we show that while deep convolutional neural networks provide good annotation accuracy, the training process can potentially introduce subtle but consistent biases. Namely, we show that unbalanced distribution of the sky location of the galaxies in the training set can lead to a consistent bias of the classifier, that can lead to a bias in the classifier based on the sky location of the galaxies it attempts to classify. When applied to very large databases typical to astronomical sky surveys, even a small bias can become statistically significant, and might even mislead potential users of data products generated by deep neural networks into false conclusions.

As shown in Table~\ref{field_differences}, the different fields have different densities in the populations of stars and galaxies, and the mean magnitude is also different. However, it is difficult to identify a specific link between the differences between the fields and the possible bias of the CNN trained by data collected from them. For instance, Datasets 1 and 2 have relatively similar characteristics of galaxy and star population and magnitude, while they also show bias based on the data they are trained with. Due to the non-intuitive ``black box'' nature of the way DCNNs work, qualitative analysis of the specific image features the DCNNs use to make the classifications is often challenging. Therefore, the performance of DCNNs is normally analyzed empirically. Since qualitative analysis of DCNNs is difficult, it is also difficult to profile the specific causes of such biases. Such bias can be driven from different atmospheric conditions at the time of imaging, background such as the presence of the galactic plane, subtle differences in the temperature of the CCD at the time of imaging, and other reasons, including combinations of multiple reasons. As mentioned above, the complex and non-intuitive nature of DCNNs makes it difficult to profile of the specific reasons that can lead to the bias.

Clearly, digital sky surveys that cover a large part of the sky have differences in the population of astronomical objects, and their magnitude or other measurements can be different in different parts of the sky \citep{chambers2016pan}. These differences can be ``learned'' by machine learning systems such that the differences between two parts of the sky lead to systematically biased results. Such bias can be driven by the different parts of the sky from which the training data were obtained. As shown here, that bias can be relatively strong, while also difficult to detect. When that bias is captured by machine learning systems, it can lead to systematic bias in the results of the algorithm when applied to data collected by sky surveys. As deep neural networks are used to turn these data into catalogs, it should be known that these catalogs can be systematically biased.

That bias can also be difficult to predict and profile. For instance, in the example of elliptical and spiral galaxies shown here, it might be expected that when limiting the magnitude of a certain galaxy population, parts of the sky where the observation is deeper would show a higher population of spiral galaxies. That is expected as better imaging allows better identification of morphological details, which consequently allows to identify more spiral galaxies. However, if the neural network is trained with elliptical galaxies taken from the observed field and spiral galaxies taken from another field, that can increase the population of elliptical galaxies detected in that part of the sky, consequently leading to a lower number of spiral galaxies. That bias is counter-intuitive to the differences between the fields. It is therefore difficult to identify and profile that bias, or provide an immediate explanation to its source unless profiling the behavior of the deep neural network that was used to make the annotations.

The examples shown in this paper are focused on specific datasets and annotation tasks, and it is reasonable to assume that many systems based on deep neural networks are not biased. However, these examples demonstrate that such systematic bias can exist, and should be taken into account when designing neural networks for annotation of astronomical images, and when using data products generated by these neural networks. Cosmic variance, different atmospheric conditions, and even different state of the hardware when training data are acquired can affect the training of an artificial neural network, and allow the network to learn non-astronomical information that can differentiate between the classes in the training set. 

Deep convolutional neural networks and other pattern recognition techniques are now common in astronomy, and can be used for a broad range of tasks. As such neural networks heavily rely on training data, it is difficult to acquire training data that evenly covers all parts of the sky under all weather conditions and status of the hardware. Therefore, when analyzing data using deep neural networks, a certain bias is expected. Deep neural network provide an excellent solution to the classification of galaxies by their morphology, and can therefore provide an effective solution for creating large catalogs of galaxy morphology. However, the use of data products produced by these networks should therefore be used with consideration of the advantages as well as the disadvantages of deep neural networks, and should be matched to tasks for which such data annotation process is scientifically sound. Large-scale analysis of such catalogs to show subtle differences in the population of certain types of objects might be a task for which results obtained by using deep neural networks should be received with caution.


\section*{Acknowledgment}

We would like to thank the two knowledgeable anonymous reviewers for the comments that helped to improve the paper. The research was supported by NSF grant AST-1903823. The Pan-STARRS1 Surveys (PS1) and the PS1 public science archive have been made possible through contributions by the Institute for Astronomy, the University of Hawaii, the Pan-STARRS Project Office, the Max-Planck Society and its participating institutes, the Max Planck Institute for Astronomy, Heidelberg and the Max Planck Institute for Extraterrestrial Physics, Garching, The Johns Hopkins University, Durham University, the University of Edinburgh, the Queen's University Belfast, the Harvard-Smithsonian Center for Astrophysics, the Las Cumbres Observatory Global Telescope Network Incorporated, the National Central University of Taiwan, the Space Telescope Science Institute, the National Aeronautics and Space Administration under Grant No. NNX08AR22G issued through the Planetary Science Division of the NASA Science Mission Directorate, the National Science Foundation Grant No. AST-1238877, the University of Maryland, Eotvos Lorand University (ELTE), the Los Alamos National Laboratory, and the Gordon and Betty Moore Foundation.

Funding for the Sloan Digital Sky Survey IV has been provided by the Alfred P. Sloan Foundation, the U.S. Department of Energy Office of Science, and the Participating Institutions. 

SDSS-IV is managed by the Astrophysical Research Consortium for the Participating Institutions of the SDSS Collaboration including the Brazilian Participation Group, the Carnegie Institution for Science, Carnegie Mellon University, Center for Astrophysics | Harvard \& Smithsonian, the Chilean Participation Group, the French Participation Group, Instituto de Astrof\'isica de Canarias, The Johns Hopkins University, Kavli Institute for the Physics and Mathematics of the Universe (IPMU) / University of Tokyo, the Korean Participation Group, Lawrence Berkeley National Laboratory, Leibniz Institut f\"ur Astrophysik Potsdam (AIP),  Max-Planck-Institut f\"ur Astronomie (MPIA Heidelberg), Max-Planck-Institut f\"ur Astrophysik (MPA Garching), Max-Planck-Institut f\"ur Extraterrestrische Physik (MPE), National Astronomical Observatories of China, New Mexico State University, New York University, University of Notre Dame, Observat\'ario Nacional / MCTI, The Ohio State University, Pennsylvania State University, Shanghai Astronomical Observatory, United Kingdom Participation Group, Universidad Nacional Aut\'onoma de M\'exico, University of Arizona, University of Colorado Boulder, University of Oxford, University of Portsmouth, University of Utah, University of Virginia, University of Washington, University of Wisconsin, Vanderbilt University, and Yale University.


\end{document}